\documentclass[aps,prb,amsmath,twocolumn,amssymb,footinbib,superscriptaddress,showpacs]{revtex4}

\usepackage{color}
\usepackage{epsfig}
\usepackage{latexsym}
\usepackage{amssymb}
\usepackage{amsmath}
\usepackage{algorithm}
\usepackage{algorithmic}
\usepackage{wrapfig}
\usepackage[apple mac]{inputenc}
\usepackage[english]{babel}
\usepackage{times}
\usepackage{latexsym}
\usepackage{fancyhdr}
\usepackage{verbatim}
\usepackage{tabularx}
\usepackage{epsfig}
\usepackage{amsmath}
\usepackage{amssymb}
\usepackage{graphicx}
\usepackage{wasysym}

\newcommand{\qed}{\hspace*{\fill}$\square$}

\newcommand{\be}{\begin{equation}}
\newcommand{\ee}{\end{equation}}

 \newcommand{\R}{\mathbf{R}}
 
 \newcommand{\N}{\mathbf{N}}

 \newcommand{\ket}[1]{|#1\rangle}
 \newcommand{\bra}[1]{\langle #1|}

\newcommand{\fD}{\mathfrak D} %
\newcommand{\fP}{\mathfrak P} %
\newcommand{\fX}{\mathfrak{X}} %
\newcommand{\fF}{\mathfrak{F}} %
\newcommand{\fM}{\mathfrak{M}} %
\newcommand{\fO}{\mathfrak{O}} %
\newcommand{\fB}{\mathfrak{B}} %
\newcommand{\fo}{\mathfrak{o}} %
\newcommand{\fQ}{\mathfrak{Q}} %

\begin{document}

\title{On Quantum Effects in a Theory of  Biological Evolution}

\author{M.~A.~Martin-Delgado}
\affiliation{Departamento de F{\'i}sica Te{\'o}rica I, Universidad
Complutense, 28040 Madrid, Spain}

\begin{abstract}
We construct a descriptive toy model that considers
quantum effects on biological evolution starting from
Chaitin's classical framework.
There are smart evolution scenarios in which a quantum world is as favorable as classical worlds for evolution to take place. However,
in more natural scenarios, the rate of evolution depends on the degree of entanglement present in quantum organisms with respect to classical organisms.
If the entanglement is maximal, classical evolution turns out to be more favorable.
\end{abstract}



\maketitle

\section{Introduction}
\label{sec:intro}
Ever since its development by Darwin \cite{darwin_1859}, the theory of evolution stands up as the landmark of fundamental knowledge in life sciences.
In this sense, it is a theory of everything that unifies all species with a common origin. The driving principle of evolution is the 'survival of the fittest'. This leads to a
common origin to all species and biological diversity. Before it, biology was conceived as static through history. After it, 
biological effects are given a dynamical framework.

As it stands, evolution is now considered the basic principle of biology and has the same character as a physical law: it 
is true as long of all pieces of experimental evidence support it. However, this does not preclude raising the fundamental question
as to why living organism evolve. This question also arises in physical laws and the underlying issue is the search for more fundamental
principles.

In a recent work, G. Chaitin \cite{chaitin_evo_11} has challenged the status of evolution and asked the question: is it possible to give a mathematical proof of evolution?
As well as why is it that living organisms evolve.

It is apparent that in addressing such  deep questions one cannot take into account all the details that are present in a living organism, whether it is highly evolved or not.
One needs to abstract the basic features and come up with a toy model in order to be able to work with it. Chaitin has followed this method and he uses a very basic 
definition of what a living organism is and a remarkable notion of a mutation. His model and insight are inspired by his earlier works on Algorithmic Information Theory (AIT)
\cite{chaitin_75, chaitin_87}.
We will refer to it as the Chaitin model and we shall describe it in Sect.~\ref{sec:classical}.

A natural and challenging problem is how to introduce quantum effects  in the classical model of Chaitin, and then try to evaluate its consequences. This is the purpose of this paper.
Related to this, an interesting questions is:

\noindent {\bf What is more favorable, to evolve in a quantum world or in a classical world?}

The answer to this question is relevant in several ways since it could shed some light to other fundamental questions:

\noindent i/ Biological evolution was formulated as a basic feature of classical living organisms for our world is classical at the macroscopic level.
However, there could have been an earlier time previous to our current 'classical era' in which quantum effects may have played a role in evolution.
Thus, was there a quantum evolution epoch before classical evolution took place?

\noindent ii/ Alternatively, there is also the possibility that classical and quantum evolution coexists at different scales. Is this possible or favorable?

A basic assumption of our quantum model for biological evolution will be the Turing barrier: a quantum computer can not compute a problem that is uncomputable for a classical computer,
i.e. for a Turing machine (TM). For example, the Turing halting problem \cite{turing_36}  is also uncomputable for a quantum Turing machine. In his famous paper on quantum simulators, Feynman's argues that this barrier is unsurmountable \cite{feynman_82} and this is the widely accepted status on these quantum limits  \cite{berstein_vazirani_97}, despite several attempts to beat the Turing barrier \cite{calude_pavlov_02,kieu_03}.
We leave for the conclusions the interesting analysis on the possible consequences of beating the Turing barrier for the quantum Chaitin model of biological evolution.

It is very deep and insightful the use of non-computability as  something positive as opposed to how it is appreciated  in more pragmatical approaches to
the foundations  of the theory of computation.
In mathematics, there is also intrinsic randomness, and Chaitin uses non-computability as a resource to have an appropriate fitness function to challenge organism to evolve,
thereby improving and becoming more advanced. This is elaborated further in the Conclusions, Sec.\ref{sec:conlcusions}.

Schr\"{o}dinger was the precursor of studying quantum effects in DNA \cite{schrodinger_44} and he thought about the possibility that mutations were originated by 
some sort of quantum fluctuations. The notion of mutation introduced here, Sec.\ref{sec:quantum}, is far more general.

When addressing the issue of quantum effects in Chaitin biological evolution,  it is crucial to bear in mind  the following fact:

\noindent i/ Complexity classes are affected by quantum effects and they are different than in the  classical case.

\noindent ii/  Computability remains the same for both quantum and classical cases (this is the Turing barrier).

\noindent Thus, as the Chaitin model is based on {\em non-computability} as a resource for driving evolution, then apparently there should not be any quantum effects.
However, the key point is that Chaitin defines an organism as a finite-size program software. Once its size $N$ is fixed, thus being finite, it is also computable, thereby
becoming a complexity problem. Thus, the way out to this apparent paradox is to realize that for finite N-size prolbems, there is no computability issue. What it is true is
that $\forall N$, it is not possible to compute the fitness functions of Chaitin based on non-computability.

The version of algorithmic complexity introduced by Kolmogorov is not prefix-free (self-delimiting programs) and does not allow to formulate halting probabilities as in
Chaitin's version of algorithmic complexity. This is why we use the latter.

This paper is organized as follows: in Sect.\ref{sec:classical} we review the classical model introduced by
Chaitin to study classical evolution scenarios using the formalism and results of AIT; in Sect.\ref{sec:quantum} the quantum
versions of organisms, mutations and fitness functions are formulated on very general grounds;
in Sect.\ref{sec:q-algorithmic_complexity} a choice of quantum algorithmic complexity has to be made and we review
the known results for entangled and separable quantum states; 
in Sect.\ref{sec:q-omega_numbers} we introduce quantum $\Omega$ numbers which play a central role in defining
mutations in a quantum world; in Sect.\ref{sec:q-evolution_scenarios} we analyze the
total evolution time and its scaling with the number of time-steps, for several quantum evolution scenarios and quantum organisms; 
Sect.\ref{sec:conlcusions} is devoted to conclusions, prospects and further explanations. Appendix \ref{sec:appendix} explains
some basic notions of AIT and in particular, prefix-free bit-strings and its coding that are necessary to compute the complexity
of quantum mutations.

\section{Classical Chaitin Model}
\label{sec:classical}

\subsection{The Model}

The fundamental notion in Chaitin model is to consider life as evolving software. This will be specified below.
To this end, let us recall some basic notions from AIT that are needed to define the model.
Let $\fX:=\{\Lambda, 0,1,00,01,10,11,000,\ldots\}$ be the set of finite strings of binary bits, with $\Lambda$ denoting the blank space symbol.
The size or number of bits is $|x|$.
The set of infinite bit-strings is denoted as $\fX^{\infty}$. A classical computer is an application $C: \fX \times \fX \rightarrow \fX$ that takes an input data $q\in \fX$ and a program
$p\in \fX$ and acts on the input to produce an output string $C(p,q)=x\in \fX$ which is the result of the computation, assuming it halts. The concrete structure and
functioning of $C$ is given by the classical Turing Machine \cite{chaitin_75, chaitin_87}. When the input data is empty, we simply write $C(p)=x$ and when the output is simply
stopping the computer with no output, we write $C(p):\text{halts}$. A universal Turing Machine (UTM) $U$ is one that can simulate the functioning of any other TM $C$.

The notion of complexity is basic in computability theory. It tells us whether a program $p\in \fX$ or input/output data $q,x\in \fX$ have a simple structure or not.
Throughout this paper, we shall be using the notion of algorithmic complexity $H(x)$ of a generic string of bits $x\in \fX$. It was studied independently by 
Solomonoff \cite{solomonoff_64},
Kolmogorov \cite{kolmogorov_65} and Chaitin \cite{chaitin_69}, and sometimes is referred to as Kolmogorov complexity. It is defined as the shortest program that can reproduce a given string $x$ in a universal
TM:
\begin{equation}
H(x):= \underset{p: U(p)=x}{\text{min} \ \  |p|}.
\label{complexity}
\end{equation}
This notion of complexity grasp the concept that the information content of a string is more related to its intrinsic computational structure rather than to its mere size.
For example, a string like $x=0101010101010101...$ may be very large, but its structure is very simple; $x=(01)^n$, for a certain integer $n$. 
The same goes for other periodic strings or structured strings. Its complexity is bounded by a constant; $H(x)<c$. On the other side
of the complexity are the random strings  $x_r$ that are those without internal structure. This is represented by a complexity $H(x_r)\geq |x_r|$, for the best thing a TM can do
is to output the same input string $x_r$.

A remark is in order. The algorithmic complexity $H(x)$ is not computable because of the existence of the Halting problem and it is defined through a optimization process.
Nevertheless, this is no obstacle to produce good and rigorous upper bounds that are enough to quantify the complexity of programs, data etc.

The classical Chaitin model is characterized by a triplet of elements $\{ \fO,\fM,\fF\}$, whose
definitions are:

\noindent i/ {\bf Living Organism $\fO$}: it is a classical program, i.e., a piece of software that can be fed in a universal Turing Machine and produce a certain
output, or just halt or even not halt. If the program $\fO$ halts, then the output is a string of classical bits $x$. In the theory of classical computation, a program
$\fO$ can also be characterized by a certain bit-string whose size is denoted as $|\fO|$. Thus, $\fD \in \fX$.

The rationale behind this choice is an abstract process that reduces an organism to pure information encoded in its DNA. The rest of the organism such as its body, 
functionalities etc are disregarded as far as being essential to evolution is concerned.
This is an oversimplification that is inherent to this toy model and so far it is necessary in order to be able to apply tools from classical information theory (AIT).

\noindent ii/ {\bf Mutation $\fM$}: it is a classical algorithm that transforms a given organism $\fO$ into a mutated organism $\fO':=\fM(\fO)$. Thus, it represents a transformation of the DNA by the action of external agents to the classical code. Thus, $\fM: \fX \rightarrow \fX$.

This notion of mutation is an algorithmic mutation as opposed to other more typical mutations called point-wise mutation that are common to population genetics studies.
What is remarkable is that an algorithmic mutation is far richer than other notions of mutations considered thus far, and in this context, it appears as the most general change that we can consider on a given living organism (classical code).

Consider the following two very different mutations acting on a $n$-string in bitwise notation $x=x_1x_2\ldots x_n\in \fX$. One is a point-wise mutation $\fP_{n_0}$ defined as
\begin{equation}
\fP_{n_0}:  x_1x_2\ldots x_{n_0} \ldots x_n  \rightarrow x_1x_2\ldots x_{n_0}\oplus1 \ldots x_n,
\end{equation}
and the other is a bit-wise mutation $\fB$
\begin{equation}
\fB:  x_1x_2\ldots x_n  \rightarrow x_1\oplus1x_2\oplus1\ldots x_n\oplus1.
\end{equation}
While $\fP_{n_0}$ represents a local change in the classical code (DNA), $\fB$ affects globlally to a all the code.  $\fP_{n_0}$ is a typical mutation in population genetics
since it is more likely to change one single base of the genetic code than multiple changes which are exponentially unlikely. On the contrary, the bit-wise mutation produces
a drastic change in the genetic code. It turns out to be useful since it may lead to a change of specie for example.
Both mutations are necessary and they find a common framework in the algorithmic treatment of evolution. They share the same amount of complexity
$H(\fP_{n_0}) \approx H(\fB) \leq c$. Therefore, having a big mutation is not penalized during the whole history of evolution.

The evolution is a process that starts with the simplest organism $\fO_1$ and it evolves towards more complex organisms $\fO_N$ after the action of 
a series of mutations $\fM_k$, $k=1,2,\ldots,N$. The algorithmic complexity $H(\fO_k)$ measures how the new successful organisms are becoming more advanced. 

\noindent It is the action of a mutation what defines the notion of time in this model and it is given by the time-step  $k$. The total evolution time would be $N$.

\noindent iii/ {\bf Fitness Function $\fF$}: this is a cost function that evaluates whether a mutated organism has improved with respect to the original. 
Thus, $\fF: \fX \rightarrow \R$.

\noindent Let $\fO_k$ be a given organism and time-step. Then, in the next step the organism is mutated to $\fO'_k:=\fM_k(\fO_k)$. The fitness function selects whether
the new organism survives or fails:
\begin{equation}
\fO_{k+1} := \begin{cases}
\fO'_k & \text{if} \quad \fF(\fO'_k) > \fF(\fO_k);\\
\fO_k & \text{if} \quad \fF(\fO'_k) \leq \fF(\fO_k).
\end{cases}
\label{fitness}
\end{equation}

Chaitin's deep insight into the problem of biological evolution is the choice of the fitness function from AIT. The idea is to see life as evolving software,
such that a living organism is tested after a mutation has occurred. The idea is to use a testing function that is an endless resource. This way, evolution
will never be exhausted, will ever go on.
In AIT there are several functions with this remarkable property that make them specially well-suited for this task: quantities that are definable but not computable.
One example is the Busy Beaver function \cite{rado_62} $\Sigma$ . Another example is Chaitin's $\Omega$ number \cite{chaitin_87,chaitin_03,calude_02} that represents the halting probability of self-delimiting TMs.

For the Busy Beaver function $\Sigma$ there are several variants which are equally good for the purposes of fitness function, that measures the rate of evolution.
For instance, $\Sigma$ can be defined as the maximum number of $1$'s output by a TM $U$ after it halts starting from a blank input data $q=\Lambda$. To work with
$\Sigma$ it is convenient to specify the maximum size $N$ that the programs $p\in \fX$ operated by $U$ and define the output as the largest integer $k\in \fX$ in binary form
that is computed after halting $U$.
Thus, a $N$-th Busy Beaver function is denoted $\Sigma_N$ and defined
\begin{equation}
\Sigma_N:= \underset{H(k)\leq N}{\text{max}\  \ k},
\label{busy_beaver}
\end{equation}
where the algorithmic complexity \eqref{complexity} is defined for programs $p$ that compute $k=U(p)$ without input and halting.
This is a well-defined function $\Sigma_N: \N \rightarrow \N$ but it is noncomputable: it grows faster than any computable function $f(N)$, $\Sigma_N > f(N)$ for sufficiently 
large $N$. Therefore, $\Sigma_N$ cannot be bounded in the form of $\Sigma_N = O(f(N))$. This is the property that makes $\Sigma_N$ a good candidate for fitness function
since it is an endless source of creativity that enable us to test a new organism, a program $\fO$, and see whether it is smarter by checking whether it can name a bigger number.
Thus, we can use \eqref{fitness} with $\fF=\Sigma_N$ and ask how the total mutation time $T_N$ behaves as N grows. Let us mention in passing that naming increasingly bigger
numbers requires lots of creativity in the form of new functions and ways to name new numbers bigger and bigger.

A more manageable and systematic choice for fitness function is Chaitin's $\Omega$ number. To define it, it is convenient to introduce the notion of universal probability
$P_U(x)$ of a given string $x\in \fX$:
\begin{equation}
P_U(x):=\sum_{p:U(p)=x} 2^{-|p|},
\label{universal}
\end{equation}
which is the probability that a program randomly drawn as a sequence of
fair coin flips $p=p_1 p_2\ldots$ will compute the string $x$. That this is a well-defined probability distribution is a central result in AIT. It relies on some
technical details: a) the programs $p$ are not arbitrary, but self-delimiting; b) convergence of the series is guaranteed by the Kraft inequality \cite{cover_thomas_06}.
A self-delimiting program is a program that knows when to stop by itself, without additional stopping symbols. It is constructed from a set of prefix-free strings of bits:
strings that are not prefix of any other string in the set (see Appendix  \ref{sec:appendix}).
In AIT, the algorithmic complexity and the universal probability of strings
are related by a Shannon type of equation:
\begin{equation}
H(x) = -\log P_U(x) + O(1).
\label{relation}
\end{equation}

The $\Omega$ number can be defined from the universal probability once we drop any reference to any particular output string:
\begin{equation}
\Omega:=\sum_{p:U(p)=\text{halts}} 2^{-|p|},
\label{omega}
\end{equation}
It is considered as the halting probability in the theory of TMs. It measures the probability that a randomly chosen program $p$ will halt when run in a UTM
that halts. Thus, it is defined on the set of prefix-free halting programs, not for arbitrary programs. Interestingly enough, Chaitin proved that universal TM exist
for self-delimiting programs. This technical condition guarantees that $0<\Omega<1$: there are always programs that halt, but not all of them will halt due to the
halting problem. Again, $\Omega$ is well-defined and noncomputable. It hosts an inexhaustible amount of knowledge and it is thus suited for a fitness function.
In short, if $\Omega$ were computable it would imply that there is no halting problem, which is false. Like $\Sigma$, it is convenient to truncate Chaitin's number 
up to programs of size $N$:
\begin{equation}
\Omega_N:=\sum_{p: |p|<N} 2^{-|p|}.
\label{omega_N}
\end{equation}
These $\Omega_N$ are lower bounds to the actual $\Omega$.
This truncation also produces an unbounded function $\Omega_N$ that reflects its non-computability.

Chaitin uses $\Omega_k$ to define an organism $\fO_k$ and a mutation $\fM_k$ at time-step $k$, as well as the fitness function $\fF$.
Namely, an organism is defined by means of  the first $N(k)$ binary digits $\omega_i$ of $\Omega_k$:
\begin{equation} 
\fo_k:=\omega_1\omega_2\ldots\omega_{N(k)}.
\label{proto_organism}
\end{equation}
To complete the construction of the organism $\fO_k$ from the proto-organism $\fo_k$, we need two more ingredients. One is to make it a self-delimiting program by including
a prefix string $1^{N(k)}0$ (see Appendix  \ref{sec:appendix}) and the other one is to prefix a program $p_\Omega$ to read the fitness of the resulting organism. Altogether, the organism looks like:
\begin{equation} 
\fO_k:=p_{\Omega} \ 1^{N(k)}0 \ \fo_k.
\label{organism}
\end{equation}
The mutation acts on the organism by trying to improve the lower bounds on $\Omega$. According to AIT, a natural move is
\begin{equation}
\fM_k: \Omega_k \longrightarrow \Omega'_k=\Omega_k + \frac{1}{2^k}.
\label{mutation}
\end{equation}
Notice that this mutation induces, in turn,  a mutation in the organism $\fO_k \rightarrow \fO'_k$ by the rules specified in its construction above.
These mutations represent challenging an organism to find a better a better lower bound of $\Omega$ which amounts to an ever increasing source of knowledge.
To this end, the fitness  function $\fF=\Omega$ is introduced as follows:
\begin{equation}
\fO_{k+1} := \begin{cases}
\fO'_k & \text{if} \quad \Omega'_k < \Omega;\\
\fO_k & \text{if} \quad \Omega'_k \geq \Omega.
\end{cases}
\label{fitness_omega}
\end{equation}
To understand this selection, notice that no truncation $\Omega_k$ can be greater that the real $\Omega$ and thus, this represents a failure. On the contrary,
if the new truncation $\Omega'_k$ is still less than  $\Omega$, we have increased our knowledge of how many programs will halt upon running a UTM ($\Omega'_k>\Omega_k$).
As Chaitin notices, this implies the use of an oracle \cite{chaitin_evo_11}.

It is possible to define a variant of the Busy Beaver function $\tilde{\Sigma}_k$  in terms of $\Omega_N$ as  the least $N$ for which the first $k$ bits of the binary
string  of $\Omega_N$ are correct. In AIT it can be proved that both Busy Beavers are approximately equal,
\begin{equation} 
\tilde{\Sigma}_N = \Sigma_{N + O(log(N))}.
\end{equation}

\subsection{Chaitin's Evolution Scenarios}

Let us denote $T_N$ the total mutation time, i.e., the number of mutations tried in order to evolve an initial organism $\fO_1$ up to a certain more fitted organism $\fO_N$.
Depending on the strategy followed by Nature, Chaitin considers three scenarios and computes the scaling of $T_N$ with $N$. In this way, one can assess which is the best 
evolutionary scenario. The results are the following:

\begin{itemize}

\item Scenario I: {\em Exhaustive Search}.

This scenario represents that there is no strategy in Nature and every possible organism is tested regardless which was the previous organism that originated it.
Thus, there is no effective application of a fitness function but Nature explores all possible codes available in the phase space. As from AIT we know that in a given
set of strings $\fX_N$ of length up to $N$ there are  $2^N-1$ strings, then the order of the evolution time is

\begin{equation}
T_N = O(2^N).
\label{exhaustive_search}
\end{equation}
It takes an exponential time to reach a certain organism $\fO_N$.

\item Scenario II: {\em Intelligent Design}.

This scenario is the opposite to the previous one. Now, Nature is not dumb but assumed to be intelligent enough so as to know about AIT and this model of evolution.
The initial proto-organism is $\fo_1=0$.
The best strategy is to apply a process of interval halving to track down better lower bounds to $\Omega$ by applying mutations $\fM_k$, $k=1,2,\ldots,N$ in this increasing
order. Thus the mutation time takes of the order of $N$ trials:

\begin{equation}
T_N = O(N)
\label{intelligent_design}
\end{equation}
Thus, by selecting intelligently the order of the mutations, since we assume that Natures knows the structure of $\Omega$, then the total evolution time for an organism
grows linearly in $N$. 

\item Scenario III: {\em Cumulative Evolution at Random}.

A more natural assumption is that Nature choses randomly the mutations $\fM_k$ among the set of possible mutations. It is a random walk in the space of mutations.
Remarkably enough, the evolution time grows in between quadratic and cubic in $N$:

\begin{equation}
T_N = O(N^{2+\delta}), \quad 0<\delta<1.
\label{cumulative_evolution}
\end{equation}
Although this is worse than scenario II, it is still  a polynomial growth and far from the exponential growth of scenario I.

\end{itemize}

\section{Quantum Chaitin Model}
\label{sec:quantum}

The following definitions are we well-motivated when trying to bring concepts from Quantum Information Theory (QIT) into Chaitin's classical model.
They can be made even more general as discussed in Sect.\ref{sec:conlcusions}.

\noindent i/ {\bf Quantum Organism $\fO^q$}: it is a pure quantum state in a Hilbert space ${\cal H}$ of  infinitely countable qubits: $\fO^q:=\ket{\Psi}\in {\cal H}$.
In practice, we shall be dealing with a finite truncation to a number of qubits $N$ denoted as ${\cal H}_N$.

The meaning of this choice is motivated by the notion of classical organism as a program for a TM. Now, the quantum version is a pure state that encodes
the information of a quantum program. This is meaningful since we have adhered to an abstraction process in which a living organism is divested of everything
except its genetic code that is represented by a classical program. Thus, a quantum organism is not a form of quantum life, but represents quantum effects
in the classical code of DNA.

\noindent ii/ {\bf Quantum Mutation $\fM^q$}: it is a quantum algorithm that transforms the original quantum organism  $\fO^q$ into a mutated quantum organism $\fO^{\prime q}$:
\begin{equation}
\fM^q: \fO^q \longrightarrow \fO^{\prime q}.
\label{q-mutation}
\end{equation}

\noindent iii/ {\bf Quantum Fitness  $\fF^q$}:  it is a cost function that selects a mutated organism when it is fittest than the original.

The traditional characters of Quantum Information \cite{nielsen_chuang, rmp} Alice $A$ and Bob $B$, can be adapted to the 
 quantum evolution scenario:  Alice is the organism  before the mutation $A=\fO^q$ and Bob is the mutated organism $B=\fO^{\prime q}$. 
 Then, $\fM^q$ will success or fail depending on the fitness of the pair $(A,B)$.

In order to complete the above quantum definitions we need to specify how to choose a triplet $\{\fO^q,\fM^q,\fF^q\}$ in the quantum case. We shall follow the classical model and 
try to find a quantum version of organisms as lower bounds to some $\Omega$ number to be specify. Once this is done, the quantum notions of mutation and fitness function
will also follow. All this can be done by defining a notion of quantum algorithmic complexity.

\section{Quantum Algorithmic Complexity}
\label{sec:q-algorithmic_complexity}

The quantumness of the $\Omega$ number that we are searching for our definition of quantum organism will depend on
the notion of quantum algorithmic complexity $H_q$  that we decide to use. In fact, there are several versions of $H_q$ \cite{vitanyi_00, berthiaume_et_al_00, gacs_01, mora_briegel_05} 
and not all of them are equivalent. We shall choose the definition of Mora and Briegel \cite{mora_briegel_05} that is called network complexity
$H_{\text{net}}$ because of the following properties \cite{mora_briegel_05, mora_briegel_04,mora_briegel_06}:

\noindent a/ $H_{\text{net}}$  is a classical algorithmic complexity associated to a quantum state. It describes how many classical bits of information
are required to describe a quantum state of $N$ qubits.
Being classical, it will allow us to
compare to previous evolution rates on equal footing.

\noindent b/ $H_{\text{net}}$ has the special property that it requires an exponential number of classical bits for the description of generic quantum states.
In particular, it detects a sharp difference between multipartite entangled states and separable states.

The network complexity is a description that Alice does of a quantum state $\ket{\Psi}_N$ she has and she wants to send this information to Bob through a
classical channel so that Bob could eventually reproduce that state on his side. It describes the classical effort Bob would have to do. In order to define network
complexity, we need several operational elements: a) a universal set of quantum gates ${\cal S}$; b) an alphabet to code circuit operations ${\cal A}$, and c) a fidelity or degree of
precision $\epsilon\in (0,1)$. With the aid of these elements, we can construct a mapping from quantum states in ${\cal H}_N$ to finite strings $\fX$, such that
\begin{equation}
\fQ_{\text{cl}}: \ket{\Psi}_N \longmapsto x_{\psi},
\label{classical_application}
\end{equation}
and then, 
\begin{equation}
H_q (\ket{\Psi}) = H_{\text{net}}(\ket{\Psi}):=H(x_{\psi}).
\label{network_complexity}
\end{equation}
The first equality represents our choice of quantum algorithmic complexity while the second is the definition of network complexity \eqref{complexity}.

The mapping $\fQ_{\text{cl}}$ \eqref{classical_application} is constructed from the elements a)-c) as follows: let us select a universal finite set of gates
for example, the one generated by the gates ${\cal S}=\{U_{\text{H}}, U_{\text{K}}, U_{\text{Cnot}}\}$ \cite{boykin_99}, i.e., the Hadamard gate, the $\pi/8$-phase gate
and the Cnot gate, respectively. Then, Alice sets up a quantum circuit of gates called $U$ by concatenating gates from ${\cal S}$, and constructs a state, namely,
 $U\ket{0}_N$, from an initialization state $\ket{0}_N:=\ket{0}^{\otimes N}$. This prepared state can approximate the desired state $\ket{\Psi}_N$ with precision
 given by
\begin{equation}
{}_N\bra{\Psi}U\ket{0}_N \geq 1-\epsilon.
\label{precision}
\end{equation}
In all what follows, $\epsilon$ will be a fixed parameter once and for all from the beginning.

Next, Alice needs to use the alphabet ${\cal A}$ in order to code all the operations in the circuit $U$ and preparation of the state with $\epsilon$-precision \eqref{precision}.
This is represented by a certain string of bits ${\cal A}(U,\epsilon):=\alpha_1\alpha_2\ldots\alpha_M$, where $M$ is the length of the resulting bit-string and is a certain function
of the number of qubits $N$. Then, the mapping  \eqref{classical_application} is given by
\begin{equation}
\fQ_{\text{cl}}(\ket{\Psi}_N)=x_{\psi}:= {\cal A}(U,\epsilon).
\end{equation}
With this, the network complexity \eqref{network_complexity} is well-defined. An additional minimization process is assumed in \eqref{complexity} since the circuit $U$ is not unique and it is natural to request to use the minimal circuit that prepares the state with the desired precission.

Our choice of quantum algorithmic complexity has very important consequences for studying quantum effects in biological evolution:

\noindent 1/ According to this definition of quantum algorithmic complexity in terms of a classical network complexity, we realize that the set of quantum states is mapped onto the set of bit-strings. Thus, while the former is uncountable, the latter is infinitely denumerable. 

\noindent 2/ By virtue of this mapping we are complying with the Turing barrier.

\noindent 3/ The fact that the network complexity is classical will make that our quantum $\Omega^q$ will be also real numbers and not quantum states or operators.
However, we can make classical definitions of $\Omega$ numbers that represent different types of quantum states (see later).

\noindent 4/ In a traditional quantum information scenario, Bob needs to agree with Alice on which alphabet to use in order to communicate. In a quantum evolution scenario,
there is no need to agree on a common language for the description since there are not two observers, but a single organism that evolves.

We shall use the following fundamental results from network complexity and quantum states \cite{mora_briegel_05}. As a consequence of the Solovay-Kitaev theorem
\cite{solovay_95,kitaev_97,nielsen_chuang},
 the number of gates (bit-string) $M$ of the circuit needed to construct a given multipartite state $\ket{\Psi}_N$ grows exponentially with $N$ for a fixed accuracy  $\epsilon$.
 \begin{equation}
H_{\text{net}}(\ket{\Psi}_N) \lesssim  2^N \log\frac{1}{\epsilon}.
\label{solovay-kitaev}
\end{equation}

Furthermore, the network complexity quantifies very differently the complexity of separable and entangled states\cite{mora_briegel_05}:

\begin{itemize}

\item Separable States $\ket{\Psi}_{\text{S}}$:

\begin{equation}
H_{\text{net}}(\ket{\Psi}_{\text{S}}) \lesssim  N \log\frac{1}{\epsilon}.
\label{separable}
\end{equation}

\item Maximally Entangled States $\ket{\Psi}_{\text{E}}$:

\begin{equation}
H_{\text{net}}(\ket{\Psi}_{\text{S}}) \lesssim  N\ 2^N  \log\frac{1}{\epsilon}.
\label{separable}
\end{equation}

\item Generic States:

\begin{equation}
H_{\text{net}}(\ket{\Psi}) \lesssim  N\ 2^{E_s(\ket{\Psi})}  \log\frac{1}{\epsilon}.
\label{generic}
\end{equation}
where $E_s(\ket{\Psi})$  is the Schmidt measure which quantifies the degree of entanglement in the multipartite pure state \cite{eisert_briegel_01}.

\end{itemize}

The fact that separable states are less complex than entangled states means that separable states are more likely:
If we type a random bit-string at a computer, most likely it will correspond to a separable state. This raises a 
fundamental question: can we use the higher complexity of  entangled states to accelerate the rate of biological evolution?
To answer this question we need to introduce the corresponding quantum $\Omega$ numbers and different scenarios for mutations 
 evolution in which evolution will develop.

\section{Quantum Omega Numbers}
\label{sec:q-omega_numbers}

In order to describe different types of quantum organisms we need to define different types
of $\Omega$ numbers associated to quantum states,. Thus, we shall use the basic results on network complexity $H_{\text{net}}$.
However, we can define Omega numbers associated to selected classes of states. As we know from the geometry of the Hilbert space
of states that the set of separable states does not intersect the set of truly entangled (maximally) states, we can define $\Omega$ numbers
by restricting the sum on the programs originated by the mapping \eqref{classical_application} to those yielding either separable or entangled states.
By construction, these sets are discrete since we are using a discrete set of universal quantum gates ${\cal S}$.

\begin{itemize}

\item Separable $\tilde{\Omega}_{\text{S}}$ number:

\begin{equation}
\tilde{\Omega}^{\text{S}}:=\sum_{p_{\text{S}}:U(p_{\text{S}})=\text{halts}} 2^{-|p_{\text{S}}|},
\label{omega_separable}
\end{equation}
where $p_{\text{S}}$ is a program that describes the network complexity of a separable state $\ket{\Psi}_{\text{S}}$.
To do this sum,  we construct
all possilble separable states and apply the mapping \eqref{classical_application}
to perform the sum. As the method is constructive, the separable states are obtained on demand.

\item Entangled (maximally)  $\tilde{\Omega}_{\text{E}}$ number:
\begin{equation}
\tilde{\Omega}^{\text{E}}:=\sum_{p_{\text{E}}:U(p_{\text{E}})=\text{halts}} 2^{-|p_{\text{E}}|},
\label{omega_entangled}
\end{equation}
where $p_{\text{E}}$ is a program that describes the network complexity of a maximally entangled state $\ket{\Psi}_{\text{E}}$.
To do this sum, we fix the accuracy $\epsilon$ which behaves as an overhead factor, then we construct
all posilble maximally entangled states and apply the mapping \eqref{classical_application}
to perform the sum. The decision problem of whether a given constructed state is maximally entangled is solved by
computing its Schmidt measure and testing that it is maximal. We take this as an operational definition of maximally
entangled state in this context.

\end{itemize}

In both sums, the programs $p_{\text{S}}$ and $p_{\text{E}}$ are assumed to be prefix-free in order to guarantee their convergence.
The typical behaviour of their general terms are $2^N$ and $2^{N2^N}$, repectively. We drop off the overhead factor from now on.
From the viewpoint of AIT, we may use another equivalent definitions in terms of the network complexity explicitly:
\begin{equation}
\Omega^{\text{S}}:=\sum_{x^{\text{S}}:U(x^{\text{S}})=\text{halts}} 2^{-H_{\text{net}}(\ket{\Psi}^{\text{S}})},
\label{omega_separable2}
\end{equation}
\begin{equation}
\Omega^{\text{E}}:=\sum_{x^\text{E}:U(x^{\text{E}})=\text{halts}} 2^{-H_{\text{net}}(\ket{\Psi}^{\text{E}})}.
\label{omega_entangled2}
\end{equation}
The above quantum $\Omega$ numbers are introduced relying on the choice of quantum algorithmic complexity
in terms of network complexity. Other choices of quantum complexity may lead to different definitions of quantum
$\Omega$ numbers that may become quantum states \cite{svozil_95,svozil_95b} or even quantum operators.

\section{Quantum Evolution Scenarios}
\label{sec:q-evolution_scenarios}
We want to compare quantum evolution in a world of maximally entangled quantum organisms w.r.t. a classical world both in  intelligent design
and  cumulative evolution scenarios.

In order to study quantum effects in evolution scenarios as in Sect.\ref{sec:classical}, \eqref{exhaustive_search} \eqref{intelligent_design} \eqref{cumulative_evolution}, we need to define a triplet  $\{\fO^q_k,\fM^q_k,\fF^q_k\}$. This is achieved 
by introducing truncated versions of the quantum $\Omega$ numbers in \eqref{omega_separable2}, \label{omega_entangled2} as follows. For separable states, we have
\begin{equation}
\Omega^{\text{S}}_N:=\sum_{n<N} 2^{-H_{\text{net}}(\ket{\Psi}^{\text{S}}_n)},
\label{omega_separable2_truncation}
\end{equation}
where the sum runs over truncations up to $N$ qubits, $\ket{\Psi}^{\text{S}}_n \in {\cal H}_n$ $n=1,2,\ldots,N-1$, 
corresponding to the construction process described in \eqref{omega_separable}, \eqref{omega_separable2}.
The quantum separable organism is a lower bound to \eqref{omega_separable2_truncation}. The key distinctive feature
is that the typical behaviour of one element in this truncated sum decreases as $2^{-k}$. Thus, the corresponding mutation
is defined such as to produce a significant change in the organisms as
\begin{equation}
\fM^q_k: \Omega^{\text{S}}_k \longrightarrow\Omega^{\prime \text{S}}_k=\Omega^{ \text{S}}_k + \frac{1}{2^k}.
\label{mutation_separable}
\end{equation}
Therefore, the analysis of the evolution rates for the quantum evolution scenarios dealing with separable states are similar
to those with classical organisms in \eqref{exhaustive_search}, \eqref{intelligent_design}, \eqref{cumulative_evolution},.
A classical state is a state that can be prepared classically, thus it can evolve classically. The same treatment 
as with the classical scenarios in Sect.\ref{sec:classical} reproduces the same evolution rates.

A different result will be obtained with maximally entangled organisms. Now, let us introduce the truncated entangled $\Omega$ number as
\begin{equation}
\Omega^{\text{E}}_N:=\sum_{n<N} 2^{-H_{\text{net}}(\ket{\Psi}^{\text{E}}_n)},
\label{omega_entangled2_truncation}
\end{equation}
This allows us to obtain a quantum version of the triplet  $\{\fO^q_k,\fM^q_k,\fF^q_k\}$. In particular, the quantum entangled organism $\fO^q_k$ at time step $k$
is defined by the same process in Sect.\ref{sec:classical},  \eqref{organism} of producing lower bounds but now with the truncated quantum $\Omega$ number \eqref{omega_entangled2_truncation}.
$\fO^q_k$ yields a lower bound to $\Omega^{\prime \text{E}}_N$ defined above \eqref{omega_entangled2_truncation}.

Next, we introduce a mutation $\fM^q_k$ that tries to make this quantum organism $\fO^q_k$ to progress. A significative progress 
will occur if we try to increase the form of the quantum $\Omega$ number \eqref{omega_entangled2_truncation} according to the
typical behaviour of its terms in the sum. This is given by $2^{-N2^N}$ for spaces up to $N$ qubits. Thus, now we define an
entangled version of the mutation as
\begin{equation}
\fM^q_k: \Omega^{\text{E}}_k \longrightarrow\Omega^{\prime \text{E}}_k=\Omega^{ \text{E}}_k + \frac{1}{2^{k2^k}}.
\label{mutation_entangled}
\end{equation}
Notice that this choice of move in the space of quantum organisms is motivated by the typical behaviour of quantum circuits
representing quantum algorithms acting on quantum states. This is the natural scale for quantum mutations to occur at the 
level of quantum organisms.

The fitness function is determined by the oracle of $\Omega^{\text{E}}$ which decides whether the mutated organism 
$\fO^{\prime q}_k$ with \eqref{mutation_entangled} succeeds or fails according to the criteria \eqref{fitness_omega}.

Now, we have all the ingredients to analyze the rates for different quantum evolution scenarios, mainly with entangled organisms.

\subsection{Quantum Exhaustive Search}
\label{sec:q-exhaustive_search}

As indicated by its name, this strategy is defined by searching all classical possible programs that can be generated quantum states available in the 
Hilbert space ${\cal H}_N$ of $N$ qubits by means of the mapping \eqref{classical_application}. For a strings of length $M$ we know this grows as
$2^M$. In turn, the length of these strings is related to the number of qubits as $M\approx 2^N$. Thus, as in this evolution scenario each mutation is exhaustive,i.e., 
it tries every possible quantum organism regardless which original organism may be, then the evolution rate behaves as
\begin{equation}
T_N = O(2^{2^N}).
\label{q-exhaustive_search}
\end{equation}

\subsection{Quantum Intelligent Design}
\label{sec:q-inteligent_design}

This strategy is like climbing a hill via the optimal path, knowing such a path before hand. In such a way that each step is always better than the previous one.
There is no backtracking. 
A more proper name would be Quantum Optimal Evolution.

Now, we have to use the quantum mutations \eqref{mutation_entangled}. 
If we produce an optimal ordered sequence of these mutations $M_k^q$ as follows: $k=1,2,...,N$ we shall reach the first N valid digits of $\Omega^{ \text{E}}_N$, by construction,
and then the evolution rate is:
\begin{equation}
T_N = O(N)
\label{q-intelligent_design}
\end{equation}
Thus, quantum intelligent design behaves linear in the number of trials $N$ in a maximally entangled world. 
This behaviour is equal as the intelligent design in a classical world \eqref{intelligent_design}.
Notice that the quantum mutations have a different  growth rate than classical mutations, but nevertheless the evolution time is the same: they are optimal.

\subsection{Quantum Cumulative Evolution}
\label{sec:q-cumulative_evolution}

This strategy is like climbing a hill, but now we do not have a priori knowledge of the best strategy to improve the lower bounds
of the quantum  number. Thus, a natural strategy is to mutate by means of a  random walk in the space of quantum mutations given by \eqref{mutation_entangled}.
In this case, the quantum mutations must be drawn at random and often enough so as to produce the same final quantum organism.

The quantum mutation is characterized by the growth $k2^k$. For simplicity, we shall take it as the leading behaviour $2^k$.
As the we have chosen the network complexity as our measure for quantum algorithmic complexity, we can now use classical formulas and
Appendix  \ref{sec:appendix}, \eqref{self-delimiting_improved_size} to estimate the complexity of a quantum mutation associated to a maximally entangled state:
\begin{equation}
H(\fM^q_k)  \lesssim  \log2^k + 2\log\log2^k + O(1) \approx k + 2\log k.
\label{mutation_complexity}
\end{equation}
Its probability is $2^{-H(2^k)} = 1/(2^k k^2)$ and its frequency is 
$k^2 2^k.$
The total evolution time $T_N$ is of the order of 
\begin{equation}
T_N = \sum_{k=1,\ldots, N} k^2 2^k, 
\end{equation}
which grows exponentially up to polynomial factors,
\begin{equation}
T_N = O(2^N).
\end{equation}
Thus, quantum cumulative evolution in a maximally entangled world behaves exponentially worse than cumulative evolution in a classical world.
Quite likely, it is more favorable to evolve in a classical world than in a quantum world. This may explain why we live in a classical world at the macroscopic level.
We should remark that this conclusion does not contradicts the fact that quantum algorithms can be more efficient than classical algorithms since our conclusions
refer to algorithmic complexity, while quantum algorithms deal with computational complexity (time and space resources for computation).

\section{Conclusions and Prospects}
\label{sec:conlcusions}
We have studied quantum effects on biological evolution by means of a descriptive toy model based on quantum algorithmic complexity.
This is an adequate option when studying biological evolution from a broad perspective and in a very large time scale, so large that any
type of quantum mutation \eqref{q-mutation} can take place and not just point-wise mutation that only affect a base in the DNA code.
In quantum evolution terms, the quantum complexity is a measure of how difficult has been for Nature to 'prepare' the quantum organism.
The results obtained in Sect.\ref{sec:q-evolution_scenarios} for the rates in quantum evolution scenarios are based not on the notion of runtime complexity,  
but on the notion of mutation time, as well as  what a typical quantum mutation move is.

The halting problem and other noncomputable functions are preceded by an aura of being a pathology, a nuisance ... eventually, something negative. This is the perspective of non-specialists.
On the contrary, we may consider this undecidability as a sort of intrinsic randomness in Mathematics. This is analogous to the intrinsic randomness that quantum theory brought
to Physics even earlier in the history of Science. Now we know from Quantum Information Theory (QIT) that this randomness can be used to our benefit, in a large variety of ways.
Similarly, It is very remarkable how Chaitin turns the problem of non-computability in algorithmic complexity into a source of creativity in order to challenge living organisms to evolve
by becoming increasingly more advanced. This process of challenging by means of mutations is endless precisely because the fitness function employed is non-computable
and cannot be bounded when truncated, as we learn from AIT. Thus, non-computability is given a positive role in a descriptive version of biological evolution.

We have adopted the same perspective when formulating a quantum version of an algorithmic model for biological evolution. This has motivated us to use 
a quantum notion of complexity based on the network complexity. In this way, We can still work with lower bounds of quantum $\Omega$ numbers as prototype
of quantum effects in DNA code. This perspective is non-trivial in the quantum case since it implicitly assumes the existence of the Turing barrier also in the quantum realm.
This is still an open problem. While a classical Turing Machine works with data and programs that are infinite but countable, a quantum Turing Machine works with non-countable
sets like complex numbers. Thus, we could argue that  the classical halting prolbem does not apply since now the number of quantum TMs is uncountable.
However, let us recall that Turing's halting problem is just one instance, very remarkable, of G\"{o}del's incompleteness theorem. Thus, if we believe that G\"{o}del's incompleteness theorem applies beyond Arithmetic, we may accept that there are uncomputable problems in quantum TMs also,  and likely something equivalent to a quantum Turing halting problem. 
Moreover, in our case, we have employed a finite set of universal quantum gates and a transformation to bit-strings from network complexity.
This implies that we are not considering a quantum TM as a continuous system, but we are dealing effectively with a countable set of gates up to a fixed precision $\epsilon$. T
his seems the simplest generalization of the classical scenario.

It is interesting to realize that the Turing barrier has important consequences in a quantum evolution scenario of this kind. In case that barrier could be beaten by quantum effects,
that would imply that we cannot use real quantum non-computability as a source of creativity in quantum evolution as in the classical Chaitin model. We could not justify quantum
effects on biological evolution on the same theoretical grounds.

The evolution rates in quantum scenarios are understood up to an overhead factor arising from the accuracy factor that we want to use. This is fixed and thus removed
from the expressions for simplicity. However,
this $\epsilon$ parameter is new in the quantum evolution case and does not exist in the classical case. 
Something similar could be introduced in a classical evolution by invoking the existence of classical errors during evolution, but this is not standard in AIT.
The reason for the existence of $\epsilon$ is because the universal gate set  ${\cal S}$ is finite. With a continuous universal gate set \cite{barenco_et_al_95,nielsen_chuang,rmp}
it is possible to get rid of it, but that would imply that Nature would had an infinite amount of resources, something which we do not consider  reasonable. 
The fundamental origin of this difference is the fact that the set of classical strings is countable while the set of quantum states is uncountable. Thus, working with a classical universal set of gates does not  need an $\epsilon$ parameter.
In this regard, the quantum complexity is more 'natural' than the classical where something like $\epsilon$ is absent at the very fundamental level.
In other words, the classical universal Turing machine and the finite universal quantum gate set are not on equal footing, but the quantum case is more 'natural' since Nature can also make errors.

Alternatively, we can think of this parameter as a grid or lattice spacing but in the space of quantum states, rather than in real space. It is a  discretization. In this sense, we always work with a finite lattice or grid,  and that is why we drop this dependence. We never take the continuum limit.

The network complexity is formulated in terms of a finite universal gate set  ${\cal S}$, instead of a quantum Turing Machine which would seem more natural if we see how
the classical algorithmic complexity is defined explicitly in terms of a classical Turing Machine. However, this is not an obstacle since we are using the Solovay-Kitaev theorem
to reconstruct arbitrary quantum unitary gate to a given precision. Furthermore, we also know that the quantum circuit model is equivalent to the quantum Turing Machine
model due to the Yao theorem \cite{yao_93}. Moreover, we have also identified that the choice of a coding language in network complexity to transform a quantum circuit
in a bit-string \eqref{classical_application} is irrelevant for quantum evolution, since Alice and Bob are replaced by the original organism and the mutated organism, respectively.

The simple quantum toy model of Sec.\ref{sec:quantum} can be thought of as a first step towards more realistic models and it does not exhaust all necessary ingredients to describe
quantum effects in biological evolution, even from a algorithmic information viewpoint. For example, we can mention a series of extensions that this model still allows:

\noindent {\em Fitness Functions}: instead of using lower bounds to quantum $\Omega$ numbers, there are other options considered by Chaitin in his classical model
that it is worthwhile to find its instance in the quantum model.

\noindent {\em Creation of Hierarchies}: classical evolution favors the formation of hierarchies of living organisms. Are they compatible with quantum evolutions? 

\noindent {\em Mixed States}: in all our analysis, quantum organisms have been represented by means of pure quantum states and mutations by quantum algorithms.
There is a natural extension to mixed quantum states  where the lack of purity here may represent the action of an external environment on the organism, i.e., its genetic code.
This degree of freedom may influence the evolution rates and it is also a way to model the presence of errors during evolution.

\noindent {\em Continuous Variables (CV)}:  we may also choose the Hilbert space of states not to be represented by qubits, but for continuous variable states \cite{CV}, 
like Gaussian states.
This is still a well-defined model for quantum computation and it remains a challenge to study its properties from the algorithmic complexity perspective and in the context of evolution.

\noindent {\em Quantum Complexity}: as we have mentioned, there are other notions of quantum algorithmic complexity that are not equivalent to network complexity.
It is still possible to keep a version of this toy model in Sect.\ref{sec:quantum} and investigate the consequences in quantum evolution of these other choices. 
In particular, quantum $\Omega$ numbers can be replaced by quantum states or even quantum operators. This may affect the evolution rates.
However, it is important to justify conceptually these other choices.

\appendix

\section{Size of Self-Delimiting Bit-Strings in Algorithmic Complexity Theory (AIT)}
\label{sec:appendix}

The size of an integer $k\in\N$ is defined as
\begin{equation}
\text{size}(k):=1 + \lceil \log(1+x)\rceil,
\label{integer}
\end{equation}
where $x\in\fX$ is the finite bit-string representation of $k$.

In AIT, it is technically necessary to work with self-delimiting, i.e., prefix-free strings
of bits in order to have a well-defined halting probability $\Omega$ that be convergent.
Let $x$ denote a $n$-bit string: $x=x_1x_2...x_n$.
The set of $x\in\fX$ strings is not self-delimiting.
For example, for  $n=2$ then $\fX=\{0,1,00,01,10,11\}$ has a 0 that is a prefix of 00 and 01 and so on.

Given a set of strings $\fX$ we can make them into a set of self-delimiting strings
by the following procedure:
\begin{equation}
 x  \longmapsto 0^n 1 \ x =: \fo_x
 \label{self-delimiting}
 \end{equation}
i.e., we put $n$ 0s before the string and use a 1 to separate it from the given $n$-bit string $x$. In the context of evolution, this is called a proto-organism.
As the Turing Machine has to read the input string bit by bit, then this way we are telling the TM the length of the string $x$ ahead of time (before the TM reads it).
Another example for $n=3$ bit-strings is
\begin{equation}
\begin{matrix}
 0 \mapsto 010 ,& 00\mapsto 00100,  & 10\mapsto 00110 , \\
1 \mapsto  011, & 01\mapsto 00101. & 11\mapsto  00111,
\end{matrix}
\end{equation}
and now we do not have prefix strings anymore.

The size of self-delimiting strings enters in the definition of the Chaitin numbers and we need to compute its size.
Denote  $|x|$ the size in bits of a string $x\in\fX$. In our case, $|x|=n.$
Then, the size of its self-delimiting extension \eqref{self-delimiting} is
\begin{equation}
|\fo_x| = 2 n +1.
 \label{self-delimiting_size}
 \end{equation}
However, we realize that the coding of the size of the string $x$ in the above self-delimiting procedure is highly inefficient 
since we are using the unary code. We can improve this coding by using  the fact that an integer like 
$|x|$  can be coded with $\log n$ bits, for $x$ large \eqref{integer}.
Thus, let us define a better encoded self-delimiting string $x'$ as follows
\begin{equation}
x':=|\fo_x|_{\text{red}} \ x,  
 \label{self-delimiting_improved}
 \end{equation}
where $|\fo_x|_{\text{red}}$ here is the string of bits representing the $\log$ of the size of $\fo_x$,  and appears before $x$.
Now, its size is  \eqref{self-delimiting_size}
\begin{equation}
|x'| =  n + 2\log n +1.
\label{self-delimiting_improved_size}
 \end{equation}

\begin{acknowledgments}

M.A.M.-D. thanks the Spanish MICINN grant FIS2009-10061,
CAM research consortium QUITEMAD S2009-ESP-1594, European Commission
PICC: FP7 2007-2013, Grant No.~249958, UCM-BS grant GICC-910758.

\end{acknowledgments}


\end{document}